\documentclass[a4paper,twoside,12pt]{article}
\input{obsjkt.sty}

\newcommand{\etal}{\textit{et al.}}                               

\newcommand{\targ}{TRAPPIST-1}

\usepackage{rotating}

\begin{document} 

\OBSheader{Transit timing measures for \targ}{J.\ Southworth \etal}{2022 Aug}

\OBStitle{VLT, GROND and Danish Telescope observations of transits in the \targ\ system}

\OBSauth{John Southworth$^1$, L.\ Mancini,$^{2,3,4}$ M.\ Dominik,$^5$ U.\ G.\ J{\o}rgensen$^6$, \\
V.\ Bozza$^{7,8}$, M.\ J.\ Burgdorf$^{\,9}$, R.\ Figuera Jaimes$^{10}$, L.\ K.\ Haikala$^{11}$, \\
Th.\ Henning$^3$, T.\ C.\ Hinse$^{12,13}$, M.\ Hundertmark$^{14}$, P.\ Longa-Pe\~na$^{15}$, \\
M.\ Rabus$^{16}$, S.\ Rahvar$^{17}$, S.\ Sajadian$^{18}$, J.\ Skottfelt$^{19}$ and C.\ Snodgrass$^{20}$}

\OBSinst{Astrophysics Group, Keele University, Staffordshire, UK}
\OBSinst{University of Rome ``Tor Vergata'', Rome, Italy}
\OBSinst{Max Planck Institute for Astronomy, Heidelberg, Germany}
\OBSinst{INAF –- Turin Astrophysical Observatory, Pino Torinese, Italy}
\OBSinst{University of St Andrews, St Andrews, UK}
\OBSinst{Centre for ExoLife Sciences, Niels Bohr Institute, Copenhagen, Denmark}
\OBSinst{Universit{\`a} di Salerno, Fisciano, Italy}
\OBSinst{Istituto Nazionale di Fisica Nucleare, Sezione di Napoli, Napoli, Italy}
\OBSinst{Universit{\"a}t Hamburg, Hamburg, Germany}
\OBSinst{Universidad San Sebastian, Valdivia, Chile}
\OBSinst{Universidad de Atacama, Copiapo, Chile}
\OBSinst{Nicolaus Copernicus University, Toru{\'n}, Poland}
\OBSinst{Chungnam National University, Daejeon, South Korea}
\OBSinst{Zentrum f{\"u}r Astronomie der Universit{\"a}t Heidelberg, Germany}
\OBSinst{Centro de Astronom\'{\i}a, Universidad de Antofagasta, Chile}
\OBSinst{Universidad Cat\'olica de la Sant\'{\i}sima Concepci\'on, Concepci\'on, Chile}
\OBSinst{Department of Physics, Sharif University of Technology, Tehran, Iran}
\OBSinst{Department of Physics, Isfahan University of Technology, Isfahan, Iran}
\OBSinst{The Open University, Milton Keynes, UK}
\OBSinst{University of Edinburgh, Royal Observatory, Edinburgh, UK}


\OBSabstract{\targ\ is an ultra-cool dwarf that hosts seven known transiting planets. We present photometry of the system obtained using three telescopes at ESO La Silla (the Danish 1.54~m telescope and the 2.2~m MPI telescope) and Paranal (Unit Telescope 1 of the Very Large Telescope). We obtained 18 light curves from the Danish telescope, eight from the 2.2~m and four from the VLT. From these we measure 25 times of mid-transit for four of the planets (b, c, f, g). These light curves and times of mid-transit will be useful in determining the masses and radii of the planets, which show variations in their transit times due to gravitational interactions.}



\section*{Introduction}

\targ\ is an ultra-cool dwarf of mass $0.089 \pm 0.006$~M$_\odot$, radius $0.121 \pm 0.003$~R$_\odot$, and effective temperature $2516 \pm 41$~K \cite{Vangrootel+18apj}. Two transiting planets were found in this system by Gillon \etal\ \cite{Gillon+16nat}, based on photometry from the 0.6-m TRAPPIST telescope \cite{Jehin+11msngr}. A further five transiting planets were found by Gillon \etal\ \cite{Gillon+17nat} and Luger \etal\ \cite{Luger+17natas} from observations with the {\it Spitzer} and K2 satellites augmented by ground-based data.

Based on analysis of transit light curves, the radii of the planets have been found to range from 0.76~R$_\oplus$ to 1.13~R$_\oplus$ and their orbital periods from 1.5~d to 18.8~d ~\cite{Delrez+18mn,Grimm+18aa,Agol+21psj}. All seven planets are in orbital resonances with each other \cite{Grimm+18aa,Agol+21psj,Teyssandier++22aa}, causing dynamical interactions between the planets which depend on their masses and orbital characteristics. This has allowed measurement of their masses, which range from 0.33~M$_\oplus$ to 1.37~M$_\oplus$ \cite{Grimm+18aa,Agol+21psj}, from analysis of transit timing variations (e.g.\ Refs.\ \cite{HolmanMurray05sci,Agol+05mn}).

\targ\ remains the lowest-mass stellar object known to host a transiting planet\footnote{Based on data obtained from the Transiting Extrasolar Planet Catalogue (TEPCat \cite{Me11mn} at \texttt{https://www.astro.keele.ac.uk/jkt/tepcat/}) on 2022/05/11.}, so the system is an important one for further study. Aside from mass and radius measurements, it is an important tracer of tidal effects \cite{HayMatsuyama19apj,Brasser++19mn,Bolmont+20aa}, the formation and interior structure of rocky planets \cite{Coleman+19aa,Turbet+20aa,Burn+21aa,Raymond+22natas}, and the characterisation of atmospheres via transmission spectroscopy \cite{Ducrot+18aj,Burdanov+19mn,Krishnamurthy+21aj,Gressier+22aa}. \targ\ is a high-priority target for observations with the James Webb Space Telescope \cite{BarstowIrwin16mn,Morley+17apj,Bean+18pasp,Krissansen+18aj,Lustig+19aj}.

The faintness of the \targ\ system ($V = 18.8$, $I = 14.0$) and the low planet masses means that it is difficult to measure the masses of the planets using high-precision spectroscopic radial velocities. Therefore measurements of the times of mid-transit for these planets are crucial for improving measurements of their masses, and thus densities and surface gravities. In this work we present extensive photometry of transits of \targ\ obtained in the 2017 and 2018 observing seasons using three telescopes.


\section*{Observations with the Danish telescope}

A total of 18 light curves of \targ\ were obtained in 2017 June--August using the 1.54\,m Danish Telescope at ESO La Silla, Chile, equipped with the DFOSC imager. The data were obtained whilst the telescope was being operated by the MiNDSTEp Consortium \cite{Dominik+10an} in the context of the transit project running as a side project in this consortium \cite{Me+09mn,Me+09mn2}. A Cousins $I$ filter was used for all observations of \targ, which is an extremely red star. A total of 13 transits were detected, with others lost to poor weather or the shifting of the transit outside the observing interval due to dynamical effects.

Data reduction was performed using the {\sc defot} pipeline \citep{Me+09mn,Me+14mn}, which uses an IDL implementation of the {\sc daophot} aperture photometry routine \cite{Stetson87pasp}. The observations were taken in focus and were debiassed and flat-fielded. Differential photometry versus multiple comparison stars was obtained by optimising the weights of the comparison stars simultaneously with a low-order polynomial to minimise the scatter around zero differential magnitude outside transit.

\begin{figure}[h] \centering \includegraphics[width=\textwidth]{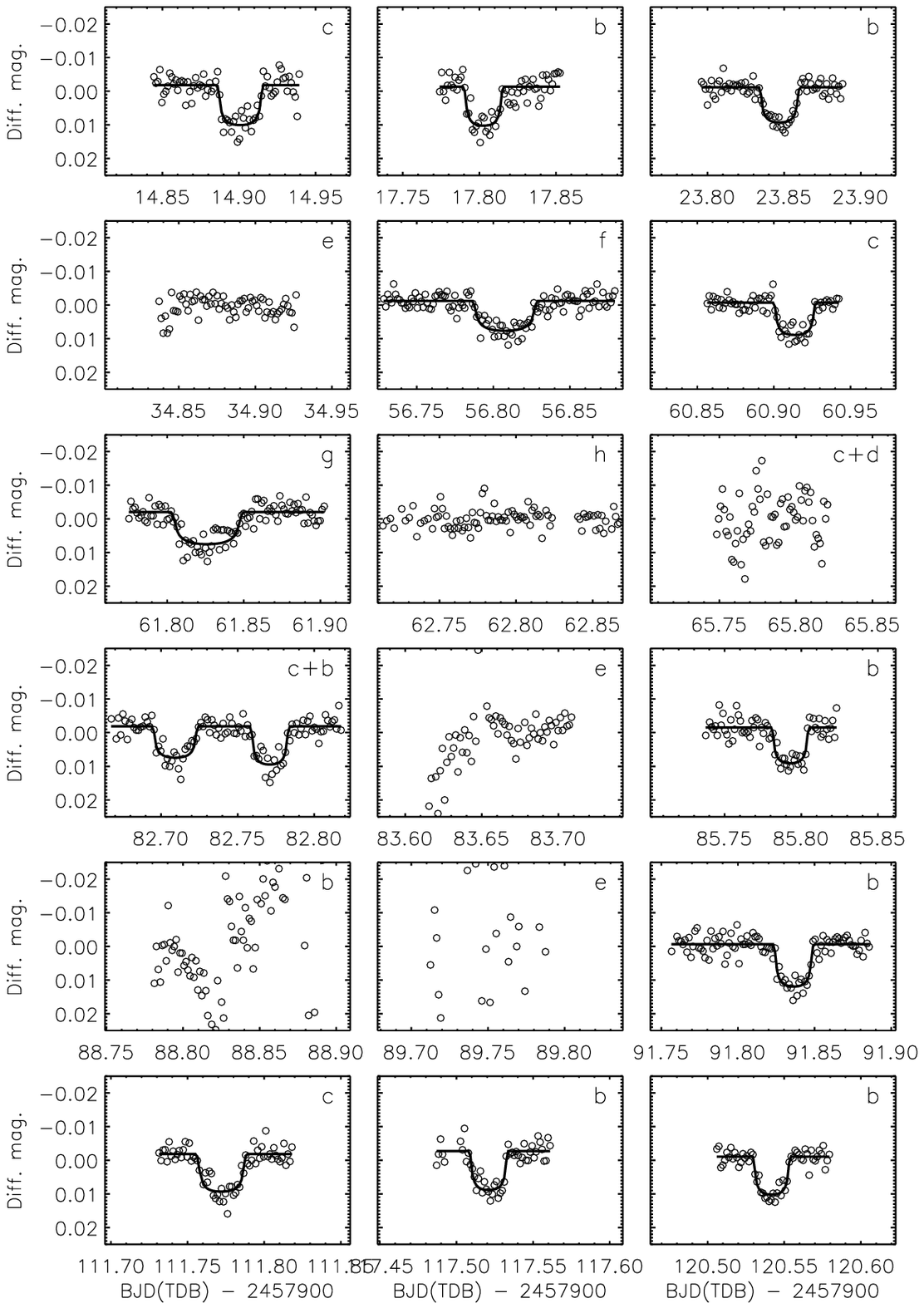} \\
\caption{\label{fig:dk} Plot of the light curves of \targ\ from the Danish Telescope.
Each panel shows one light curve on the same scale (0.16\,d and 0.05\,mag). The data
are shown using open circles and the {\sc jktebop} best fits using solid lines. For
clarity, the errorbars are not plotted. The designation(s) of the planet(s)
transiting are shown at the top-right of each panel.} \end{figure}

An observing log is given in Table~\ref{tab:dk}. The light curves are shown in Fig.~\ref{fig:dk}. The timestamps have been placed on the BJD(TDB) timescale using routines from Eastman \etal\ \cite{Eastman++10pasp}. The times written into the FITS files were manually checked during the observation of the majority of the transits to confirm their reliability.

\begin{sidewaystable}
\begin{center}
\caption{\em \label{tab:dk} Log of the observations from the Danish telescope. $N_{\rm obs}$ is the
number of observations, $T_{\rm exp}$ is the exposure time, $T_{\rm dead}$ is the dead time between
exposures, `Moon illum.' is the fractional illumination of the Moon at the midpoint of the transit,
and $N_{\rm poly}$ is the order of the polynomial fitted to normalise the light curve to zero differential
magnitude. The aperture radii refer to the target aperture, inner sky and outer sky, respectively.}
\vskip 1mm
\setlength{\tabcolsep}{4.5pt}
\begin{tabular}{llcccccccccccr}
{\em Telescope} & {\em Planet(s)} & {\em Date of}   & {\em Start} & {\em End}   & {\em Filter} & $N_{\rm obs}$ & $T_{\rm exp}$ & $T_{\rm dead}$ & {\em Airmass} & {\em Moon}   & {\em Aperture}   & $N_{\rm poly}$ & {\em Scatter} \\
                &                 & {\em first obs} & {\em (UT)}  & {\em (UT)}  &              &               &   {\em (s)}   &   {\em (s)}    &               & {\em illum.} & {\em radii (px)} &                & {\em (mmag)}  \\[8pt]
Danish   & c         & 2017/06/10 & 08:13 & 10:30 &  $I$   &      74    &   100       &       12      & 1.32 $\to$ 1.10            & 0.876 &  9 \,14 \,30 &      1       &  2.9  \\
Danish   & b         & 2017/06/13 & 06:31 & 08:24 &  $I$   &      63    &   100       &        8      & 1.96 $\to$ 1.25            & 0.992 &  8 \,12 \,30 &      1       &  2.9  \\
Danish   & b         & 2017/06/19 & 07:02 & 09:15 &  $I$   &      72    &   100       &       11      & 1.50 $\to$ 1.12            & 0.301 &  8 \,13 \,30 &      1       &  1.9  \\
Danish   & e $^*$    & 2017/06/30 & 08:00 & 10:09 &  $I$   &      67    &   100       &       16      & 1.15 $\to$ 1.10 $\to$ 1.13 & 0.432 &  9 \,13 \,30 &              &  2.8  \\
Danish   & f         & 2017/07/22 & 05:21 & 08:58 &  $I$   &     116    &   100       &       11      & 1.50 $\to$ 1.12            & 0.019 &  7 \,12 \,25 &      1       &  2.1  \\
Danish   & c         & 2017/07/26 & 08:25 & 10:28 &  $I$   &      67    &   100       &       12      & 1.13 $\to$ 1.51            & 0.117 &  8 \,12 \,25 &      1       &  1.8  \\
Danish   & g         & 2017/07/27 & 06:27 & 10.01 &  $I$   &      98    &   100       &       13      & 1.13 $\to$ 1.10 $\to$ 1.44 & 0.187 &  6 \,12 \,25 &      1       &  2.6  \\
Danish   & h $^*$    & 2017/07/28 & 04:50 & 08:52 &  $I$   &      92    &   200--100  &       12      & 1.38 $\to$ 1.10 $\to$ 1.22 & 0.272 &  7 \,13 \,25 &              &  3.7  \\
Danish   & c, d $^*$ & 2017/07/31 & 04:25 & 07:32 &  $I$   &      88    &   200--100  &        6      & 1.44 $\to$ 1.10            & 0.561 &  4 \,~8 \,20 &              &  6.0  \\
Danish   & c, b      & 2017/08/17 & 03:51 & 08:27 &  $I$   &     103    &   100       &       25      & 1.49 $\to$ 1.10 $\to$ 1.17 & 0.258 &  8 \,12 \,25 &      1       &  2.2  \\
Danish   & e $^*$    & 2017/08/18 & 02:36 & 04:50 &  $I$   &      69    &   100       &       25      & 1.72 $\to$ 1.14            & 0.170 &  6 \,10 \,20 &              &  3.1  \\
Danish   & b         & 2017/08/20 & 05:32 & 07:25 &  $I$   &      68    &   100       &        8      & 1.10 $\to$ 1.21            & 0.030 &  9 \,13 \,25 &      1       &  2.6  \\
Danish   & b $^*$    & 2017/08/23 & 06:34 & 09:12 &  $I$   &      86    &   100       &       11      & 1.13 $\to$ 1.77            & 0.030 &  6 \,11 \,25 &              &  8.2  \\
Danish   & e $^*$    & 2017/08/24 & 04:55 & 07:14 &  $I$   &      84    &   100       &       16      & 1.11 $\to$ 1.10 $\to$ 1.25 & 0.075 &  5 \,~8 \,20 &              & 40.1  \\
Danish   & b         & 2017/08/26 & 05:59 & 09:05 &  $I$   &      98    &   100       &       13      & 1.11 $\to$ 1.80            & 0.224 &  9 \,13 \,25 &      1       &  2.4  \\
Danish   & c         & 2017/09/15 & 05:19 & 07:27 &  $I$   &      66    &   100       &       13      & 1.15 $\to$ 1.64            & 0.279 &  9 \,13 \,25 &      2       &  2.4  \\
Danish   & b         & 2017/09/20 & 23:29 & 01:17 &  $I$   &      51    &   100       &       11      & 2.42 $\to$ 1.37            & 0.008 &  8 \,13 \,25 &      1       &  2.5  \\
Danish   & b         & 2017/09/23 & 23:58 & 01:44 &  $I$   &      57    &   100       &       13      & 1.83 $\to$ 1.23            & 0.149 &  8 \,12 \,25 &      1       &  1.9  \\
\end{tabular}
\end{center}
$^*$ transit not detected due to observing conditions or dynamical effects.
\end{sidewaystable}

\section*{Observations with the MPI 2.2~m telescope}

\targ\ was observed on four night using the MPI 2.2~m telescope at ESO La Silla. The GROND imager \cite{Greiner+08pasp} was used to obtain observations simultaneously in seven passbands, approximating the Gunn $g$, $r$, $i$ and $z$ and near-infrared $J$, $H$ and $K$ filters. The $g$ and $r$ bands suffer from a high scatter due to the faintness of the target. Conversely, the $J$, $H$ and $K$ bands do not yield good light curves because the target is much brighter than the available comparison stars. We therefore present light curves from only the $i$ and $z$ bands here.

We encountered a problem with the presence of transient bad pixels in the $r$ and $z$ bands caused by a bug in the CCD controller software. These bad pixels comprise approximately 0.5\% of the pixels in an image and cause the pixel count rate to be either the bias level or unusually high. Left uncorrected, the bad pixels significantly increase the scatter in the light curves. Fortunately, the bad pixels occurred randomly in each image. We were therefore able to find them by comparing each pixel value in two successive images to detect those that differed by more than a manually-chosen threshold. The count rates in the affected pixels were then replaced by the means of the count rates in the surrounding eight pixels.

\begin{figure}[t] \centering \includegraphics[width=\textwidth]{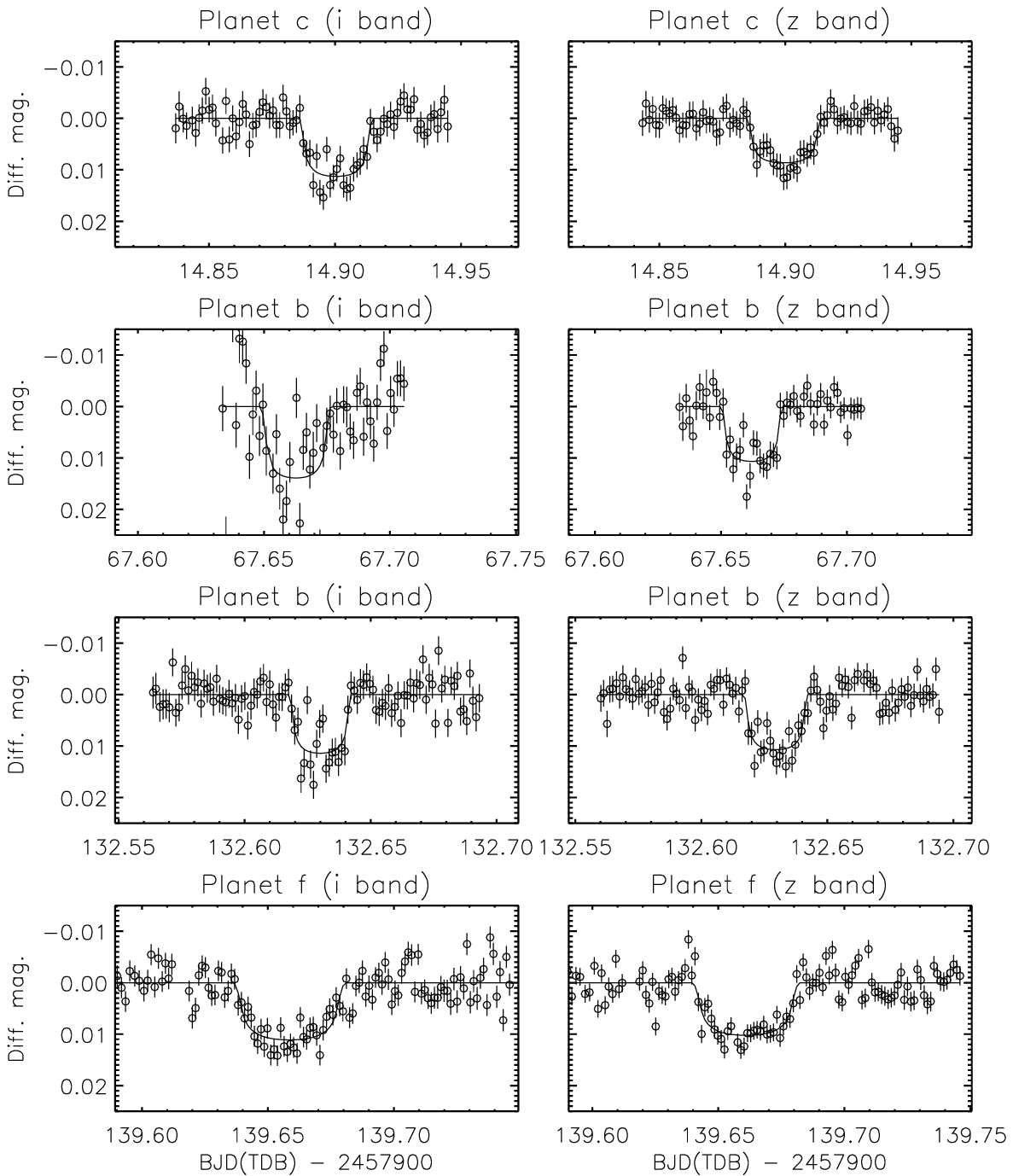} \\
\caption{\label{fig:grond} Plot of the light curves of \targ\ from the MPI 2.2~m
telescope. The data are shown using open circles and the {\sc jktebop} best fits
using solid lines. Panels on the same row show data from the same transit.} \end{figure}

Aside from the bad-pixel correction, the data were reduced using the {\sc defot} pipeline as described above. The observing log is given in Table~\ref{tab:grond+vlt} and the data are plotted in Fig.~\ref{fig:grond}.

\begin{sidewaystable}
\begin{center}
\caption{\em \label{tab:grond+vlt} Log of the observations from the MPI 2.2~m telescope and the VLT. Other comments are as in Table~\ref{tab:dk}.}
\vskip 1mm
\setlength{\tabcolsep}{4.5pt}
\begin{tabular}{llcccccccccccr}
{\em Telescope} & {\em Planet(s)} & {\em Date of}   & {\em Start} & {\em End}   & {\em Filter} & $N_{\rm obs}$ & $T_{\rm exp}$ & $T_{\rm dead}$ & {\em Airmass} & {\em Moon}   & {\em Aperture}   & $N_{\rm poly}$ & {\em Scatter} \\
                &                 & {\em first obs} & {\em (UT)}  & {\em (UT)}  &              &               &   {\em (s)}   &   {\em (s)}    &               & {\em illum.} & {\em radii (px)} &                & {\em (mmag)}  \\[8pt]
GROND    & c         & 2017/06/10 & 07:49 & 10:37 & $i$               &      80    &    90       &       25      & 1.40 $\to$ 1.10            & 0.876 & 15 \,22 \,40 &      2       &  2.4  \\
GROND    & c         & 2017/06/10 & 07:49 & 10:37 & $z$               &      77    &    90       &       25      & 1.40 $\to$ 1.10            & 0.876 & 20 \,28 \,50 &      2       &  1.6  \\
GROND    & b         & 2017/08/02 & 03:01 & 04:49 & $i$               &      54    &    90       &       25      & 2.17 $\to$ 1.31            & 0.747 & 27 \,35 \,60 &      2       &  8.6  \\
GROND    & b         & 2017/08/02 & 03:01 & 04:49 & $z$               &      55    &    90       &       25      & 2.17 $\to$ 1.31            & 0.747 & 34 \,40 \,70 &      2       &  2.7  \\
GROND    & b         & 2017/10/06 & 01:17 & 04:30 & $i$               &     105    &    60--80   &       27      & 1.19 $\to$ 1.10 $\to$ 1.21 & 0.997 & 14 \,23 \,50 &      2       &  3.1  \\
GROND    & b         & 2017/10/06 & 01:17 & 04:30 & $z$               &     108    &    60--80   &       27      & 1.19 $\to$ 1.10 $\to$ 1.21 & 0.997 & 18 \,28 \,60 &      1       &  2.6  \\
GROND    & f         & 2017/10/13 & 01:41 & 05:45 & $i$               &     119    &    80       &       31      & 1.11 $\to$ 1.10 $\to$ 1.70 & 0.429 & 16 \,25 \,50 &      2       &  3.1  \\
GROND    & f         & 2017/10/13 & 01:41 & 05:45 & $z$               &     118    &    80       &       31      & 1.11 $\to$ 1.10 $\to$ 1.70 & 0.429 & 20 \,30 \,55 &      2       &  2.7  \\
VLT      & c         & 2017/11/05 & 02:23 & 04:29 & $z_{\rm special}$ &     192    &    12--8    &       26      & 1.18 $\to$ 1.77            & 0.980 & 10 \,15 \,35 &      1       &  3.1  \\
VLT      & c         & 2018/09/06 & 05:45 & 07:52 & $z_{\rm special}$ &     230    &    10--6    &       26      & 1.13 $\to$ 1.55            & 0.167 & 11 \,20 \,40 &      1       &  2.4  \\
VLT      & c         & 2018/09/11 & 02:13 & 04:17 & 815 / 13          &     145    &    25       &       26      & 1.31 $\to$ 1.09            & 0.028 & 12 \,20 \,40 &      1       &  2.3  \\
VLT      & b         & 2018/11/14 & 23:56 & 01:54 & $z_{\rm special}$ &     195    &    25--6    &       25      & 1.10 $\to$ 1.21            & 0.447 & 10 \,30 \,40 &      1       &  2.8  \\
\end{tabular}
\end{center}
\end{sidewaystable}

\section*{Observations with the VLT}

One transit of \targ~b and three more of \targ~c were observed using the Very Large Telescope (VLT) Unit Telescope 1 (UT1) equipped with the FORS2 imager \cite{Appenzeller+98msngr}. The data were obtained through either a $z_{\rm special}$ filter with a central wavelength of 916~nm and a full-width at half maximum of 18~nm, or a night-sky suppression filter with a central wavelength of 813~nm and a full-width at half maximum of 13~nm. These observations were obtained in order to search for a variation in radius with wavelength indicative of the presence of a planetary atmosphere, as tentatively found for GJ~1132 \cite{Me+17aj}, but this test could not be performed due to the scatter of the observations plus complications with the scheduling of these time-critical observations in service mode.

\begin{figure}[t] \centering \includegraphics[width=\textwidth]{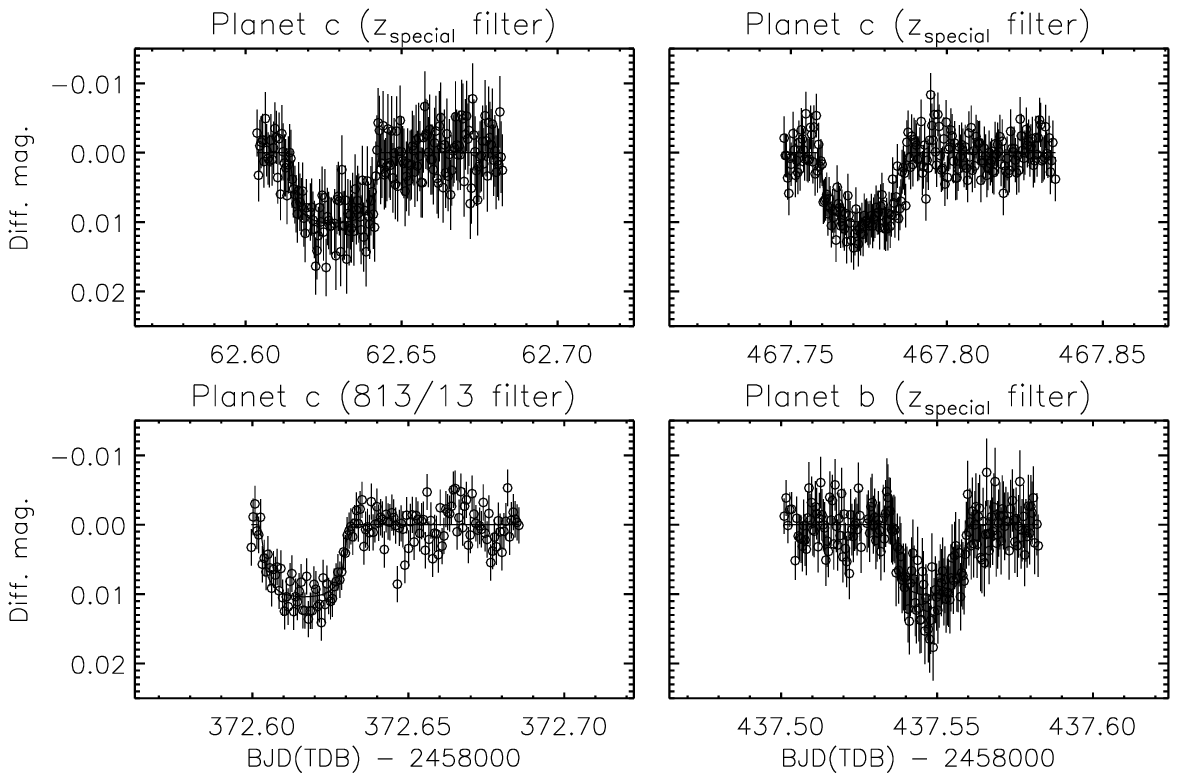} \\
\caption{\label{fig:vlt} Plot of the light curves of \targ\ from the VLT. The data are
shown using open circles and the {\sc jktebop} best fits using solid lines.} \end{figure}

The data were reduced using the {\sc defot} pipeline as described above. The observing log is given in Table~\ref{tab:grond+vlt} and the data are plotted in Fig.~\ref{fig:vlt}. These data have a relatively large scatter because all comparison stars had significantly lower count rates than the target star.

\section*{Transit timing measurements}

Each light curve was modelled individually using version 38 of the {\sc jktebop}\footnote{\texttt{http://www.astro.keele.ac.uk/jkt/codes/jktebop.html}} code \cite{Me++04mn2,Me13aa}. We fitted for the sum of the fractional radii ($r_{\rm A}+r_{\rm b}$ where $r_{\rm A} = \frac{R_{\rm A}}{a}$, $r_{\rm b} = \frac{R_{\rm b}}{a}$, $R_{\rm A}$ is the radius of the star, $R_{\rm b}$ is the radius of the planet in question, and $a$ is the semimajor axis of the relative orbit), their ratio ($k = \frac{r_{\rm b}}{r_{\rm A}}$) and the time of transit mid-point. We fixed the orbital periods and inclinations to the values measured by Gillon \etal\ \cite{Gillon+17nat}. Limb darkening was accounted for using the quadratic law with coefficients fixed at $u=0.25$ and $v=0.60$.

The light curve from the Danish telescope on the night of 2017/08/17 contains two transits, the first by planet c and the second by planet b. These were modelled separately after removing the data for the other transit from the light curve before fitting.

Four of our transit observations were obtained simultaneously in the $i$ and $z$ passbands using GROND. This provides an opportunity to check if the two light curves from each night yield timings in mutual agreement. We calculated  the level of agreement to be 1.5$\sigma$ for the night of 2017/06/10, 0.3$\sigma$ for 2017/08/02, 0.8$\sigma$ for 2017/10/06 and 2.9$\sigma$ for 2017/10/13. This agreement is acceptable but suggests that the errorbars of our timing measurements may be slightly too small.

The resulting transit times are given in Table~\ref{tab:tmin}. The reduced light curves will be made available at the \textit{Centre de Donn\'ees astronomiques de Strasbourg\footnote{\texttt{http://cdsweb.u-strasbg.fr/}}}. Because all planets in the \targ\ system show complex transit timing variations caused by gravitational interactions, and because good ephemerides are available from elsewhere (Refs.\ \cite{Gillon+17nat,Delrez+18mn,Agol+21psj}), we have not performed any analysis on the transit timings in the current work. They are instead presented here so they may be used in future studies of this system.

\begin{table}
\begin{center}
\caption{\em \label{tab:tmin} Times of mid-transit measured for the planets in the TRAPPIST-1 system.}
\begin{tabular}{lccc}
{\em Telescope} & {\em Filter} & {\em Planet} & {\em Time of mid-transit (BJD/TDB)} \\[5pt]
GROND  & $i$               & c      & 2457914.90035 $\pm$ 0.00039 \\
GROND  & $z$               & c      & 2457914.89956 $\pm$ 0.00034 \\
Danish & $I$               & c      & 2457914.90084 $\pm$ 0.00047 \\
Danish & $I$               & b      & 2457917.80266 $\pm$ 0.00049 \\
Danish & $I$               & b      & 2457923.84702 $\pm$ 0.00042 \\
Danish & $I$               & f      & 2457956.80753 $\pm$ 0.00054 \\
Danish & $I$               & c      & 2457960.91366 $\pm$ 0.00036 \\
Danish & $I$               & g      & 2457961.82616 $\pm$ 0.00067 \\
GROND  & $i$               & b      & 2457967.66265 $\pm$ 0.00098 \\
GROND  & $z$               & b      & 2457967.66231 $\pm$ 0.00042 \\
Danish & $I$               & c      & 2457982.70928 $\pm$ 0.00051 \\
Danish & $I$               & b      & 2457982.77099 $\pm$ 0.00039 \\
Danish & $I$               & b      & 2457985.79287 $\pm$ 0.00042 \\
Danish & $I$               & b      & 2457991.83630 $\pm$ 0.00032 \\
Danish & $I$               & c      & 2458011.77209 $\pm$ 0.00046 \\
Danish & $I$               & b      & 2458017.52100 $\pm$ 0.00039 \\
Danish & $I$               & b      & 2458020.54178 $\pm$ 0.00030 \\
GROND  & $i$               & b      & 2458032.63003 $\pm$ 0.00044 \\
GROND  & $z$               & b      & 2458032.62955 $\pm$ 0.00038 \\
GROND  & $i$               & f      & 2458039.65854 $\pm$ 0.00064 \\
GROND  & $z$               & f      & 2458039.66100 $\pm$ 0.00057 \\
VLT    & $z_{\rm special}$ & c      & 2458062.62830 $\pm$ 0.00031 \\
VLT    & $z_{\rm special}$ & c      & 2458367.77302 $\pm$ 0.00023 \\
VLT    & 815 / 13          & c      & 2458372.61662 $\pm$ 0.00027 \\
VLT    & $z_{\rm special}$ & b      & 2458437.54799 $\pm$ 0.00028 \\
\end{tabular}
\end{center}
\end{table}

\section*{Acknowledgements}

This paper was based on data collected by MiNDSTEp with the Danish 1.54\,m telescope at the ESO La Silla Observatory. This paper was also based on observations collected using the Gamma Ray Burst Optical and Near-Infrared Detector (GROND) instrument at the MPG 2.2~m telescope located at ESO La Silla, Chile, under programs 099.A-9030(A) and 0100.A-9004(A). GROND was built by the high-energy group of MPE in collaboration with the LSW Tautenburg and ESO, and is operated as a PI-instrument at the MPG 2.2~m telescope. This paper was also based on observations collected at the European Southern Observatory under ESO programme 0100.C-0716(C). The following resources were used in the course of this work: the NASA Astrophysics Data System; the SIMBAD database operated at CDS, Strasbourg, France; and the ar$\chi$iv scientific paper preprint service operated by Cornell University.
We thank Eric Agol for checking our planet identifications and alerting us to one incorrect one.

UGJ acknowledges funding from the Novo Nordisk Foundation Interdisciplinary Synergy Programme grant no. NNF19OC0057374 and from the European Union H2020-MSCA-ITN-2019 under Grant no. 860470 (CHAMELEON).

NP's work was supported by Funda\c{c}\~ao para a Ci\^{e}ncia e a Tecnologia (FCT) through the research grants UIDB/04434/2020 and UIDP/04434/2020.

PLP was partly funded by Programa de Iniciaci\'on en Investigaci\'on-Universidad de Antofagasta, INI-17-03.


\bibliographystyle{obsmaga}

\end{document}